\documentclass[%
 reprint,%
 amssymb, amsmath,%
twocolumn
]{revtex4-1}

\usepackage{bm}%
\usepackage[colorlinks=true,linkcolor=blue]{hyperref}%

\usepackage{graphicx}
\usepackage{amsmath}
\usepackage{epstopdf}
\usepackage{color}

\expandafter\ifx\csname package@font\endcsname\relax\else
 \expandafter\expandafter
 \expandafter\usepackage
 \expandafter\expandafter
 \expandafter{\csname package@font\endcsname}%
\fi
\begin{document}

\title{Ground state phase diagram of ultrasoft bosons}%

\author{Alejandro Mendoza-Coto}%
\affiliation{Departamento de F\'\i sica, Universidade Federal de Santa Catarina, 88040-900 Florian\'opolis, Brazil}%

\author{Diogo de Souza Caetano}%
\affiliation{Departamento de F\'\i sica, Universidade Federal de Santa Catarina, 88040-900 Florian\'opolis, Brazil}%

\author{Rogelio D\'iaz-M\'endez}%
\affiliation{Department of Physics, KTH -- Royal Institute of Technology, SE-10691 Stockholm, Sweden}%

\begin{abstract}
	In 2D bosonic systems ultra-soft interactions develop an interesting phenomenology that ultimately leads to the appearance of supersolid phases in free space conditions.
	While suggested in early theoretical works and despite many further analytical efforts, the appearance of these exotic phases as well as the detailed shape of the ground-state phase diagrams have not been established yet.      
	Here we develop a variational mean-field calculation for a generic quantum system with cluster-forming interactions.
	We show that by including the restriction of a fixed integer number of particles per cluster the ground-state phase diagram can be obtained in great detail.
	The latter includes the determination of coexistence regimes of crystals of different occupancy as well as crystals with superfluid phases.
	To illustrate the application of the method we consider 
	the softened van der Waals potential, for which the phase diagram is known via quantum Monte Carlo simulations. 
	For densities other than the corresponding to the single-particle crystal, our results show very good quantitative agreement with the simulations regarding the location of the superfluid transitions.
	Additionally, the phase diagrams suggest that the solid-superfluid coexistence could be a reliable marker to locate supersolidity.
\end{abstract}

\maketitle

\section{Introduction}

The theoretical study of the quantum phase transition occurring in the crystallization of soft-core bosons have received a particular attention over the last decade~\cite{to2014,cinti14,prestipino18,heinonen19,Fa2019}.  
This interest is twofold. 
On the one hand ultrasoft bosons allow to understand exotic many-body effects in which quantum fluctuations weakens crystalline order to give rise to, e.g., supersolid states~\cite{svistunov15, pitaevskii16, panas19, prestipino19}. 
On the other hand, the advances in experimental Bose-Einstein condensation techniques involving cold atomic gases have boosted and fed back the need for theoretical insights on how to control these phases of matter~\cite{anderson95, kadau16,leonard17,chomaz18}.

In the supersolid phase, the systems still display a modulated density that breaks the rotational and translational symmetries, while at the same time develop superfluid transport properties.
The characterization of the properties of these ground-state phase diagrams using specific interaction potentials have been a fruitful direction for numerical research~\cite{AnMePu16,DiMe15,MaMepu19}.  
Several analytical studies have focused as well on the supersolid and superfluid  phases of systems with standard non-cluster-forming soft interaction potentials.~\cite{pomeau94,henkel10,saccani12,kunimi12,macri13,ancilotto13}
Properties related to, e.g., the supersolid-superfluid transition~\cite{pomeau94,henkel10} and the excitation spectra~\cite{macri13,ancilotto13} are of great current interest for the scientific community.    
Many calculations use mean-field variants~\cite{kunimi12,macri13} and more sophisticated approaches~\cite{pomeau94,henkel10} to tackle the problem in this scenario.
In this context, the ground-state phase diagrams of ultra-soft bosons, in which the crystalline structures are formed by clusters instead of single particles, is a relatively new problem that is generating theoretical and experimental advances.~\cite{cinti14,svistunov15,prestipino19}

Bosons interacting via ultrasoft potentials crystallize in a lattice of clusters~\cite{likos07,ale21} which, at very small temperatures, melts due to the quantum fluctuations produced by the zero point motion.
This quantum transition to a superfluid phase is always present in the ground state for the regime of small pressure and weak interaction strength.
Near the melting point, however, a supersolid behavior have been predicted via numerical simulations and analytical approximations~\cite{cinti14,ZhMaPo2019}.

Recently, some important efforts have also been done to analytically determine these ground-state properties in the ultra-soft cluster-forming scenario, through formulations that can be applied to different potentials.
In particular, by using generalizations of the Gross-Pitaevskii mean-field model~\cite{heinonen19} and variational formulations of the condensate wave function~\cite{prestipino18,prestipino19}, a number of predictions have been put forward.  
These predictions include the order of the transitions and the dispersion relations of the supersolid lattice.   
Other important properties like the location of the phase boundaries are determined only approximately~\cite{prestipino18,heinonen19} and not in full agreement with the outcomes of Monte Carlo simulations.   

In the present work we go a step further in the analytical understanding of the ground-state phase diagrams of ultrasoft bosons.  
By using a variational mean-field calculation we improve the predictions on the  location
 of the  phase boundaries of generic ultrasoft potentials.
The main assumption used in the present formalism is the restriction of the number of particles within the clusters to be a constant integer value.
We show that, for the model potential considered in Ref.~\cite{cinti14}, the ground-state phases and the coexistence regimes can be calculated to a degree of reliability that is consistent with Monte Carlo numerical simulations in moderate and high density. 
This interaction form is of relevance for ultracold atoms~\cite{henkel10, maucher11} and has been used to model a large variety of other soft-matter systems~\cite{mladek06,dm19,sciortino13}.

We determine the solid-superfluid coexistence region (SSC) by 
looking at the chemical and mechanical equilibrium condition that exist between the pure solid and superfluid thermodynamical phases.
Clearly, this regime can not be interpreted as a supersolid by any means.
However, for the model potential used here as example, there is a coincidence regarding the shape and the location of the SSC region analytically calculated and the supersolid phase that have been observed via exact numerical simulations~\cite{cinti14}.

\section{Method}
Consider a system of interacting bosons in 2D with a pair potential $V(\vec{r})=Uv(\vec{r})$. 
The units of energy and length are such that the kinetic energy operator of a particle is given by $\hat{T}=-(\vec{\nabla})^2/2$. 
Under these conditions the Hamiltonian of the system is given by
\begin{eqnarray}
 \hat{H}&=&\sum_i\hat{T}_i+\sum_{i<j}V(\vec{r}_i-\vec{r}_j)\\
 &=&-\frac{1}{2}\sum_i(\vec{\nabla}_i)^2+U\sum_{i<j}v(\vec{r}_i-\vec{r}_j).
\end{eqnarray}
Here the potential $v(r)$ is bounded and its Fourier transform  has a negative minimum at some finite wave vector.
These conditions determine the ultrasoft character of the interactions and ensure the cluster-forming character of system in the in the classical regime.
The characterization of the ground-state phase diagram is done by means of a variational approach. 
It is thus expected that our results will better describe the system in the regime where the interparticle distance is is small than the length scale of the potential.
A variational ground state function is proposed consistent with the mean-field approximation~\cite{prestipino18,prestipino19}
\begin{equation}
 \vert\psi\rangle=\prod_i\phi_0(\vec{x}_i),
 \label{fvar}
\end{equation}
where the single particle wave function is
\begin{equation}
\phi_0(\vec{x})=\frac{\sum_{j}c_{j}\cos(\vec{k}_j\cdot\vec{x})}{\sqrt{A(c_0^2+\frac{1}{2}\sum_{j\neq0}c_{j}^2)}}.
\end{equation}
Here $A$ stands for the area of the system and the set of vectors $\vec{k}_{j}$ are selected in such a way that the field $\phi_0(\vec{x})$ reproduce the symmetries of a triangular lattice of particles.
While other structures could be stable for specific potentials, we focus  here on triangular lattices as this is the most common structure observed in simulations of ultrasoft potentials~\cite{prestipino14,cinti14,kroiss19}.  
Consequently  $\vec{k}_{j}=(p\vec{u}_1+q\vec{u}_2)k_0$, with $p$ and $q$ integers, $u_1=(0,1)$ and $u_2=(\sqrt{3}/2,-1/2)$. 
We set $c_0=1$ without loss of generality.

These Ansatz for the single-particle ground state function are usually used in the literature to model states in which particles form a triangular crystal or even triangular crystal of clusters~\cite{prestipino18,zhang19}.
A central constraint, that in the case of cluster-crystals has not been taken into account to the best of our knowledge, is the relation between $k_0$, the lattice spacing $a$ and the occupancy number $n$ of the clusters in the crystal, which is a positive integer. 
To consider $n$ as a positive integer does not rule out states with fractional average occupation, since these states can be seen as the result of the phase coexistence of solids with consecutive integer occupation number.

Considering our definition of $\phi_0(\vec{x})$, it can be shown that for this function to be periodic with a spatial period $a$, the main modulation wave vector should be $k_0=4\pi/(\sqrt{3}a)$. 
At the same time the average density $\rho$ is related to the lattice spacing $a$ and the occupancy number $n$ by the relation 
\begin{equation}
 \rho=\frac{2n}{\sqrt{3}a^2}.
\end{equation}
This relation implies that cluster-crystals with different occupancy number, at the same density, have different lattice spacing.
As a consequence
\begin{equation}
 k_0(n,\rho)=\frac{4\pi}{\sqrt{3}}\sqrt{\frac{\sqrt{3}\rho}{2n}}.
 \label{cons}
\end{equation}
This relation is at the core of numerous properties of the cluster-forming particle system. 
The disregard of this constraint, in the assumption that $n$ can be treated as a variational real parameter, explains absence of transitions between different cluster-crystal phases in mean-field approaches when describing the zero and finite temperature properties of classical and quantum particles systems~\cite{prestipino18,dunkel19}. 

In numerical simulations cluster-forming particle systems have been observed to remain with a fixed occupancy number while density is increased~\cite{Pres2014,Pres2015}. 
While this fact have been already observed, it is generally considered that the presence of hopping between neighboring clusters justifies the consideration that $n$ and consequently $k_0$ are variational parameters~\cite{prestipino14,caprini18,wang19}.

In this work we assume that $n$ is an integer value. 
This means that the cluster-crystal phases consist in lattices of equally occupied clusters of bosons.
We use this assumption to make a complete analysis of the ground-state phase diagram indicating the regions corresponding to the different clusters phases.
At some special regions, of course, coexistence between states with occupancy number $n$ and $n+1$ exists, but only in these regions the occupancy number 
can be interpreted as varying continuously between $n$ and $n+1$. 

Proceeding with the construction of the energy functional, 
the energy per particle of the cluster crystal with $n$ particles per cluster, at an average density $\rho$, is given by
\begin{eqnarray}
 \frac{E}{N}=\epsilon_n&=&\langle\psi\vert\hat{T}_1\vert\psi\rangle+\langle\psi\vert\hat{V}_{12}\vert\psi\rangle\\ \nonumber
 &=&\frac{1}{4}\frac{\sum_{j\neq0}c_j^2k_j^2}{(1+\frac{1}{2}\sum_{j\neq0}c_{j}^2)}
 \\&+& U\rho\frac{\left(\hat{v}(0)b_0^2+\sum_{j\neq0}\hat{v}(k_j)b_j^2/2\right)}{2(1+\frac{1}{2}\sum_{j\neq0}c_{j}^2)^2},
 \label{EVq}
\end{eqnarray}
where
$k_j=\vert(p\vec{u}_1+q\vec{u}_2)\vert\frac{4\pi}{\sqrt{3}}\sqrt{\frac{2n}{\sqrt{3}\rho}}$, 
and the coefficients $b_j$ are related to $c_j$ through the relation $(\sum_{j}c_{j}\cos(\vec{k}_j\cdot\vec{x}))^2=\sum_{j}b_{j}\cos(\vec{k}_j\cdot\vec{x})$. 
The quantity $\hat{v}(k)$ stands for the Fourier transform $\hat{v}(\vec{k})$ of the potential $v(\vec{r})$,
\begin{equation}
 \hat{v}(\vec{k})=\int d^2r \ e^{-i \vec{k}\cdot\vec{r}} v(\vec{r}),
\end{equation}
once we take into account the rotation symmetry of the pair interaction potential.

By minimizing the energy per particle Eq.~(\ref{EVq}) in terms of the coefficients $c_j$ the ground-state ($U$ versus $\rho$) phase diagram of the system  can be obtained  in this mean-field approximation.

\section{Results}
For the quantum regime it has to be considered not only the different cluster configurations, characterized by its occupancy number, but also the superfluid homogeneous phase.
This homogeneous configuration can be recovered from the modulated solution 
$\phi_0(\vec{x})$ by setting $c_j=0$ for all $j\neq0$. 
So it is straightforward that the energy per particle of the homogeneous phase is given by $E_h(\rho)/N=U\hat{v}(0)\rho/2$.

To minimize analytically the function of Eq.~(\ref{EVq}) considering the whole set of coefficients $c_j$ can be a very difficult task. 
Consequently, finding closed expressions for the boundaries between the different phases is fairly complicated. 
However, the more interesting transition happening in the quantum regime, i.e. the quantum melting of the solid phases into the superfluid phase, is in fact tractable within certain approximations.
Around this transition, quantum fluctuations characterized by the amplitude of the zero point motion are expected to be strong enough to destabilize the crystalline order.
In this scenario the particles are less localized and the profile is smoother, making the Fourier amplitudes of the profile smaller for all non-zero wave vectors.
In order to analytically advance we consider the single mode approximation, $c_j=0$ for $j>1$. 
Such as an approximation can not be fully justified with qualitative arguments, and its ultimate validity can be verified by comparing with general mean-field numerical results and with previous Monte Carlo simulations.

\subsection{Analytical results on the quantum melting}
Following the single mode approximation $c_{j>1}=0$ lets proceed with the analytical study of the quantum melting of the solid phase. 
In the single mode approximation the energy per particle of the system (Eq.\ref{EVq}) yields
\begin{eqnarray}
\nonumber
 \epsilon_n&=&\frac{3k_0(n,\rho)^2c_1^2}{4(1+\frac{3}{2}c_1^2)}+\frac{U}{2}\rho\hat{v}(0)\\
 &+&\frac{3U\rho\hat{v}(k_0(n,\rho))(2c_1+c_1^2)^2}{4(1+\frac{3}{2}c_1^2)^2},
 \label{smEq}
 \end{eqnarray}
where $k_0(n,\rho)$ is given by Eq.~(\ref{cons}). 
Consequently, the difference between the energy per particle of the modulated and the homogeneous phase ($c_1=0$) is
\begin{eqnarray}
\nonumber
 \Delta\epsilon_n(c_1)&=&\frac{3k_0(n,\rho)^2c_1^2}{4(1+\frac{3}{2}c_1^2)}+\frac{3U\rho\hat{v}(k_0(n,\rho))(2c_1+c_1^2)^2}{4(1+\frac{3}{2}c_1^2)^2}.\\
 \end{eqnarray}
The boundary between each type of modulated phase with the homogeneous phase defines an energy-crossing line through the condition 
\begin{equation}
\label{syst22}
\Delta\epsilon_n(c_1^*)=0,
\end{equation}
where $c_1^*$ represent the value of $c_1$ minimizing the energy of the modulated state $\epsilon_n(c_1)$ and consequently $\Delta\epsilon_n(c_1)$, once we take into account that the energy of the homogeneous phase is independent of the value of $c_1$. Our definition for $c_1^*$ implies then
\begin{eqnarray}
\label{syst21}
 \frac{d\Delta\epsilon_n}{dc_1}(c_1^*)&=&0,
 \end{eqnarray}
Solving Eq.~(\ref{syst22}) for $c_1^*$ we concluded that:  
\begin{equation}
 c_1^*=\frac{6\tilde{U}+2\pi\sqrt{-72\pi^2+42\sqrt{3}\tilde{U}}}{12\sqrt{3}\pi^2-3\tilde{U}},
\end{equation}
where $\tilde{U}=Un\vert\hat{v}(k_0(n,\rho))\vert$ and $U$ and $\rho$ represent. 
This result can now be substituted in Eq.~(\ref{syst21}) concluding that, at the transition, 
\begin{eqnarray}
\nonumber
\tilde{U}&=&4\sqrt{3}\pi^2/7,\\
 c_1&=&1/3.
 \label{melt}
\end{eqnarray}

Eq.~(\ref{melt}) implies that, in the single mode approximation, the energy crossing line between the solid and the superfluid phase is given by the condition $\tilde{U}=4\sqrt{3}\pi^2/7$. 
This is an universal condition, independent of the occupancy number of the solid and valid for any cluster forming potential $\hat{v}(k)$. 
At the energy crossing line, the amplitude $c_1$ takes the universal value $c_1=1/3$, which can be interpret as a quantum analog of the Lindemann criterion.            
 
At this point it is possible to calculate the value of the interaction strength $U$ at the transition point from the $n$-particle cluster solid to the superfluid phase by using the first condition in Eq.~(\ref{melt}).
This leads to
\begin{equation}
 U_n(\rho)=-\frac{4\sqrt{3}\pi^2}{7n\hat{v}(k_0(n,\rho))},
 \label{Urho}
\end{equation}
where $k_0(n,\rho)$ is given by Eq.~(\ref{cons}).

For different values of $n$, Eq.~(\ref{Urho}) represents the melting curves of the different solid phases, and naturally they are valid in the region of densities in which $\hat{v}(k_0(n,\rho))$ is negative and contains the absolute minimum of $\hat{v}(k)$.
The union of the areas delimited by the curves $U_n(\rho)$ represents the whole solid region.
Interestingly, the boundary of the solid region is not a simple monotonous curve as 
obtained in mean-field calculations. 
This is in agreement with reported results of ground-state numerical simulations.\cite{cinti14} 

In order to test the single mode approximation we additionally calculated the solid-to-fluid phase boundary resulting from the minimization of the full energy function Eq.~(\ref{EVq}) for the potential $v(r)=(1+r^6)^{-1}$. 
This is a repulsive potential expressed through the dimensionless quantities $r/r_c\to r$, $v(r)/U_0\to v(r)$, where $r_c$ and $U_0$ are the units of length and energy respectively.
The potential approaches a constant value for small interparticle distances $r<1$, and drops to zero for $r\to \infty$ with a repulsive van der Waals tail~\cite{maucher11}.

\begin{figure}[ht]
	\includegraphics[width=9.cm,height=18cm,keepaspectratio]{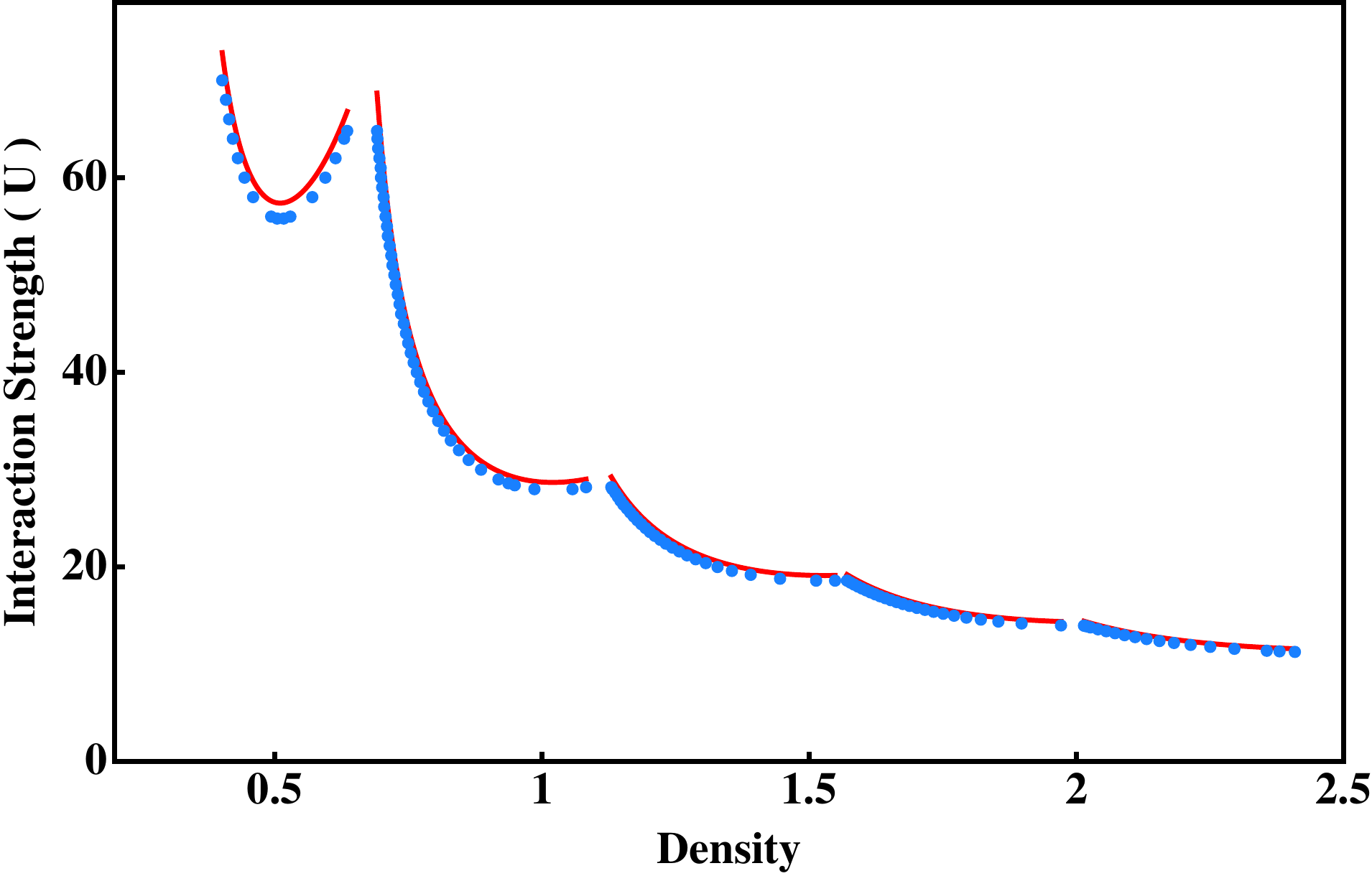}
	\caption{Comparison between numerical minimization of the full energy function Eq.~(\ref{EVq}) and the single-mode melting theory. 
	Blue dots correspond to the numerical minimization while the red curves correspond to the function $U_n(\rho)$ of Eq.~(\ref{Urho}) for $n=1,2,3,4 \mathrm{\ and\ } 5$.  
		}
	\label{fig3}
\end{figure}

In Fig.~\ref{fig3} the result from this numerical minimization (blue dots) is plotted together with the curves $U_n(\rho)$ for occupancy numbers $1\leq n\leq5$. 
As can be observed, the numerical results and the single mode melting curves compare very well, which indicates that in the melting region the single mode approximation is essentially correct. 
This is a quite useful validation because it largely simplifies all analytical calculations in the melting region.

The envelope of the family of curves $U_n(\rho)$ can be also calculated and yields
\begin{equation}
 U(\rho)=-\frac{3(2\hat{v}(k_m)-k_m^2\hat{v}''(k_m))}{14\hat{v}(k_m)\hat{v}''(k_m)\rho}.
 \label{Uenve}
\end{equation}
This envelope correspond to the result that would be obtained by considering $n$ as a real variational parameter,  which is the assumption in previous mean-field descriptions. 
Not to take into account the variation of the stability of the different cluster-crystal phases with fixed occupancy number 
can explain 
why many theoretical calculations find a monotonous behavior of the solid-to-superfluid phase boundary.
The obtained scaling $U(\rho)\propto\rho^{-1}$ of the envelope curve is also a result that have been observed with numerical simulations~\cite{cinti14}.    

\subsection{Extent of the SSC region}
For a given occupancy $n$, with varying density each cluster phase is limited by the inverted dome-shaped region of the curve $U_n(\rho)$. 
This transition from the solid to the homogeneous or superfluid state is expected to be of first order type and consequently occurs through a crossover along the coexistence region~\cite{prestipino18}.  
In the coexistence regions the pressure $P(\rho)$ and the chemical potentials $\mu(\rho)$ of each phase are equal and remain the same when density is increased from the beginning to the end of the coexistence regions. 
The mathematical condition determining the densities $\rho_{1n}$ and $\rho_{2n}$ at the beginning and end of the coexistence regions is given by   
\begin{eqnarray}
\nonumber	
 P_n(\rho_{1n})&=&P_{n+1}(\rho_{2n})\\
\mu_n(\rho_{1n})&=&\mu_{n+1}(\rho_{2n}).
\label{syst}
\end{eqnarray}
Within the canonical ensemble formalism, pressure and chemical potential can be calculated using the relations $P_n(\rho)=\rho^2\frac{\partial \epsilon_n(\rho)}{\partial \rho}$ and $\mu_n(\rho)=\epsilon_n(\rho)+\rho \frac{\partial \epsilon_n(\rho)}{\partial \rho}$~\cite{polls02}. 

To pursue the calculation of the coexistence region analytically we need to calculate the energy per particle of the cluster state as a function of the density. 
Even within the single-mode approximation, no simple analytical closed expression exist for the ground-state energy of the system. 
In order to obtain such analytical expression from Eq.~(\ref{smEq}) further approximations have to be done. 
First, the general relation between $\tilde{U}$ and the value of $c_{1m}$ minimizing de energy function Eq.~(\ref{smEq}) is found from the condition  $\frac{d\Delta\epsilon}{dc_1}(c_1)=0$. 
This leads  to the relation
\begin{equation}
 \tilde{U}=-\frac{4\pi^2(2+3c_{1m}^2)}{\sqrt{3}(-4-6c_{1m}+4c_{1m}^2+3c_{1m}^3)}.
\end{equation}
Then the value of $\tilde{U}$ is used to write a closed expression for the minimum energy per particle in terms of $c_{1m}$, eliminating the interaction strength parameter $U$.
This procedure yields
\begin{equation}
\Delta\epsilon_n(c_{1m})=\frac{\rho}{n}\frac{4\sqrt{3}\pi^2(-1+3c_{1m})c_{1m}^3}{(-4-4c_{1m}-6c_{1m}^3+9c_{1m}^4)}.
\label{inter}
\end{equation}
Now a series expansion is made for $\tilde{U}$ around the value of $c_{1m}$ at which the energies of the cluster and the homogeneous configuration equal, fulfilling Eq.~(\ref{melt}).
Inverting this expansion we can rewrite $c_{1m}$ as a power series of $(\tilde{U}-\frac{4}{7}\sqrt{3}\pi^2)$.
This expansion can be then substituted in Eq.~(\ref{inter}) to finally obtain a series expansion of the energy $\Delta\epsilon_n$ in powers of $(\tilde{U}-\frac{4}{7}\sqrt{3}\pi^2)$. 
This procedure leads to the expression
\begin{eqnarray}
\nonumber
 \Delta\epsilon_n(\tilde{U})&=&\frac{\rho}{n}\left(-\frac{1}{3}(\tilde{U}-\frac{4}{7}\sqrt{3}\pi^2)\right.\\
 &-&\left.\frac{7}{2\sqrt{3}\pi^2}(\tilde{U}-\frac{4}{7}\sqrt{3}\pi^2)^2+...\right),
 \label{Eu}
\end{eqnarray}
which is valid for $\tilde{U}\geq\frac{4}{7}\sqrt{3}\pi^2$. 
For $\tilde{U}<\frac{4}{7}\sqrt{3}\pi^2$ the minimum of $\Delta\epsilon_n$ is reached at $c_1=0$ and consequently $\Delta\epsilon_n=0$. 

Examining the system of equations in Eq.~(\ref{syst}) for the densities at the beginning and the end of the coexistence region, we realize it needs as an input the quantities $\epsilon_n(\rho)$ and its derivatives with respect to $\rho$. 
A general analytical solution of this equation system for the $n$-cluster SSC phase, for an arbitrary potential $\hat{v}(k)$, is impossible even in the single mode approximation.
We estimate it using an expansion of $\epsilon_n(\rho)$ up to first order around the density at which the energy of the cluster and fluid phases equals $\rho_0(n,U)$. 
We
proceed considering
\begin{eqnarray}
\nonumber
 \epsilon_n(\rho)&=&\frac{U\hat{v}(0)\rho_0(n,U)}{2}+\epsilon_n'(\rho_0)(\rho-\rho_0(n,U))\\
 \epsilon_h(\rho)&=&\frac{U\hat{v}(0)\rho}{2},
\end{eqnarray}
where $\epsilon_h(\rho_0)$ represents the energy of the homogeneous state. 
Additionally let us note that $\epsilon_n'(\rho_0)$ can be calculated using the obtained result for $\Delta\epsilon_n(\tilde{U})$ in Eq.~(\ref{Eu}). 
A straightforward calculation allows us to conclude that
\begin{equation}
 \epsilon_n'(\rho_0)=\frac{U\hat{v}(0)}{2}-\frac{1}{3}U\rho_0\partial_\rho\vert\hat{v}(k_0(n,\rho_0))\vert.
\end{equation}
Taking into account the above simplifications, we directly obtain the densities at the beginning and the end of the coexistence regions 
\begin{eqnarray}
\nonumber
 \rho_1(n)&=&\frac{1}{2}\rho_0(n,U)\left(1+\sqrt{\frac{\hat{v}(0)}{\hat{v}(0)-2\rho_0\partial_\rho\vert\hat{v}(k_0(n,\rho_0))\vert/3}}\right)\\ \nonumber
 \rho_2(n)&=&\frac{1}{2}\rho_0(n,U)\left(1+\sqrt{\frac{\hat{v}(0)-2\rho_0\partial_\rho\vert\hat{v}(k_0(n,\rho_0))\vert/3}{\hat{v}(0)}}\right).\\
 \label{rhoco}
\end{eqnarray}

\begin{figure}[ht]
	\includegraphics[width=9.cm,height=18cm,keepaspectratio]{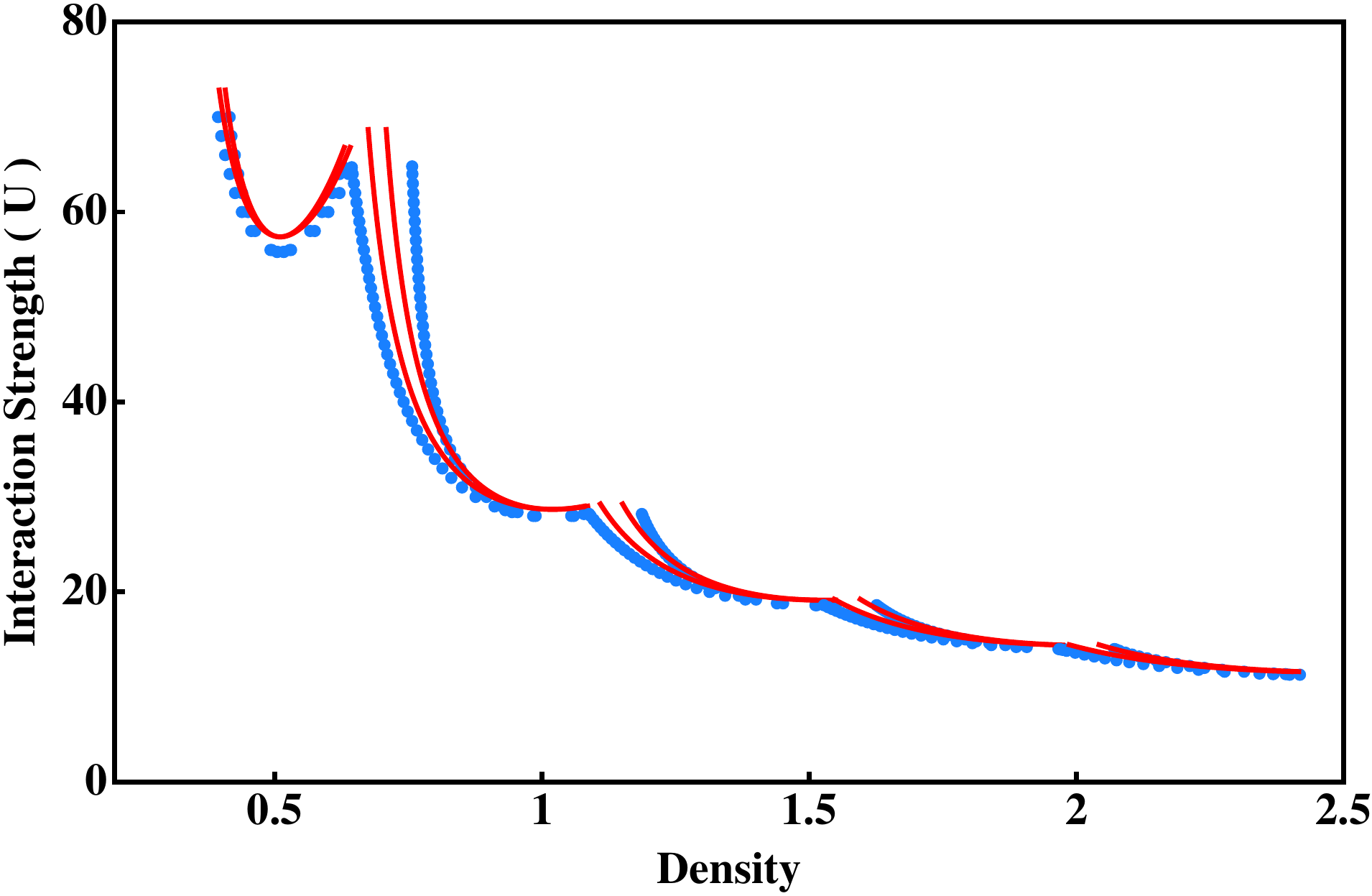}
	\caption{Coexistence regions of the solid-to-superfluid transitions. 
		Dotted curves correspond to the exact SSC determined from the minimization of the full energy function while the continuous lines represent the analytical estimate described in the main text.
		}
	\label{fig4}
\end{figure}

To asses the accuracy of the analytical estimations on the coexistence region, in Fig.~\ref{fig4} a comparison is shown between the approximate solution of Eq.~(\ref{rhoco}) and the exact values of the density at the beginning and end of the SSC. 
The exact values are determined from the exact mean-field value of $\epsilon_n(\rho)$ following the procedure described in this section. 
In Fig.~\ref{fig4} the exact coexistence regions corresponding to different occupancy numbers are represented by the blue dots while the analytical
estimation of Eq.~(\ref{rhoco}) is represented by the continuous red curves.

As can be seen, the analytical approximation underestimates the extent of the SSC. 
It is possible to show that this is a general feature resulting from neglecting higher order corrections in $\rho-\rho_0(n,U)$ for the expansion of $\epsilon_n(\rho)$. 
Nevertheless, the result serves as a simple analytical lower bound for the extent of the SSC, 
which only exists in a narrow region. 

\subsection{Numerical phase diagram}

\begin{figure}[!t]
	\hspace{-0.5cm}\includegraphics[width=9cm,keepaspectratio]{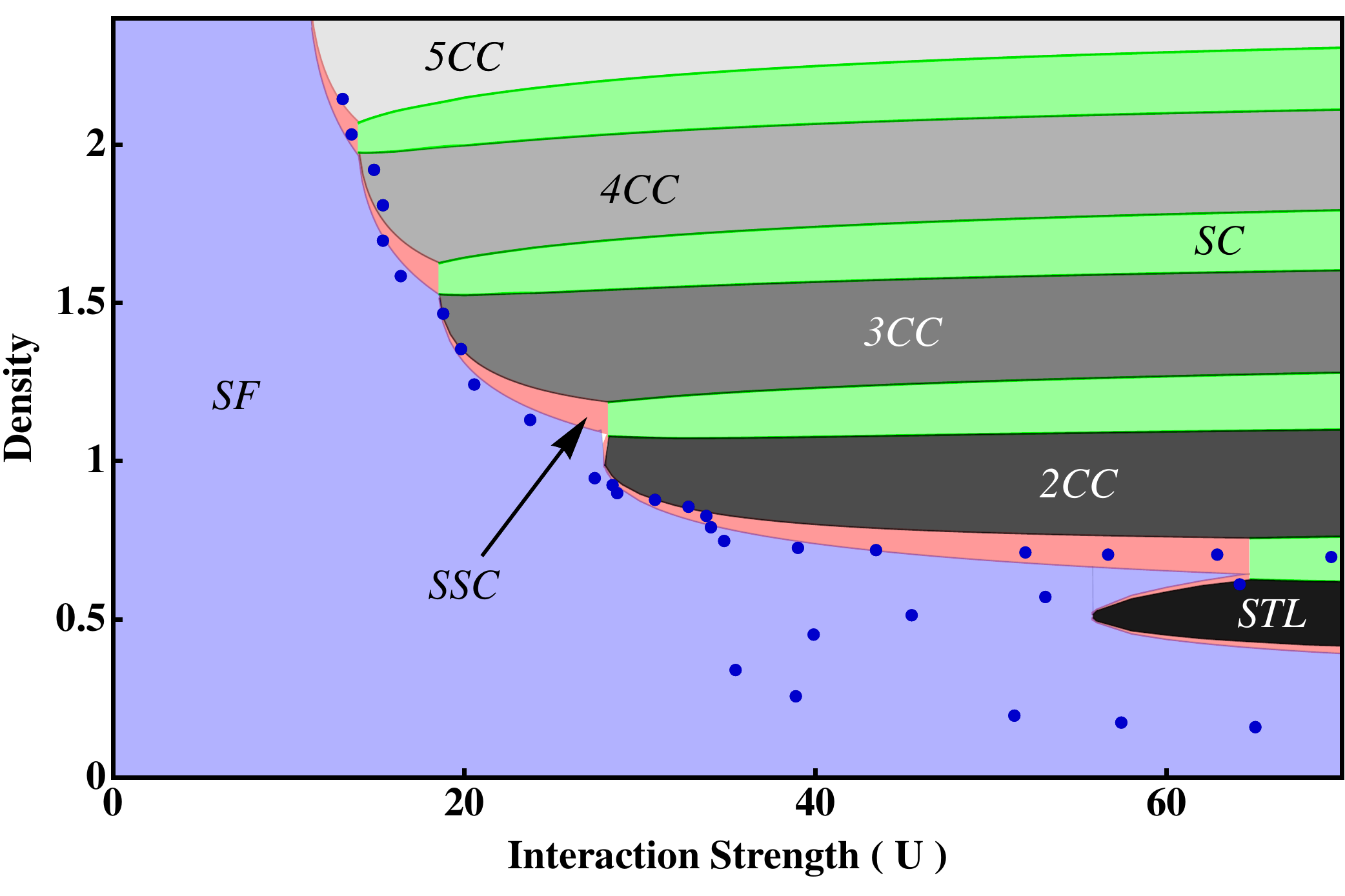}
	\caption{Ground-state phase diagram for the potential
	    $v(r)=(1+r^6)^{-1}$.
		Regions are determined by our numerical mean-field calculations.
		Pure phases are represented in light blue for the superfluid region (SF), black for the single-particle triangular lattice (STL) and gray scale for the cluster crystal solids ($n$CC). 
		The coexistence regions are colored in green for the solid coexistence (SC) and orange for the solid-superfluid-coexistence (SSC).  
		Dark blue dots are the superfluid-supersolid boundary obtained with Monte Carlo techniques in Ref.~\cite{cinti14}, taken with permission of the authors.
	}
	\label{fig2}
\end{figure}

The numerical minimization of the energy function allows us to construct the full phase diagram for the potential under consideration $v(r)=(1+r^6)^{-1}$. 
This ground-state phase diagram is shown in Fig.~\ref{fig2}, including not only the pure cluster phases ($n$CC), but also the coexistence regions between cluster solids with different occupation (SC) and the SSC.
The method followed to determine the coexistence regions in this case is the same sketched in the previous section. 
As expected, we observe that increasing density the pure cluster phase with occupancy number $n$ is limited from below by the coexistence region of the cluster transition from $n-1$ to $n$ and from above by the coexistence region of the cluster transition from $n$ to $n+1$.

As can be seen in Fig.~\ref{fig2}, the SSC only exist in a narrow region 
between the solid and the superfluid phases. 
Thus, the SSC region follows the non-monotonic behavior of the energy crossing line of the solid-to-fluid transition (see Fig.~\ref{fig3}).
This behavior is well documented with Monte Carlo simulations for the supersolid phases of this potential~\cite{cinti14}. 
Moreover, it was also reported that the density intervals corresponding to the coexistence of two solid phases are more likely to develop supersolidity, which is consistent with the phenomenology observed for the SSC in Fig.~\ref{fig2}.

For comparison, we also show in Fig.~\ref{fig2} the superfluid-supersolid boundary as numerically obtained in Ref.~\cite{cinti14} with dark blue dots.
The correspondence between the exact localization of these points and our SSC boundary without any extra fit is remarkable, although for small density this correspondence is lost.   
The superfluid phase boundary is thus quantitatively and qualitatively better described by looking at the coexistence regions, which are a direct consequence from considering that clusters phases have a fixed integer number of particles per cluster.

\begin{figure}[!t]
	\includegraphics[width=9.cm,height=18cm,keepaspectratio]{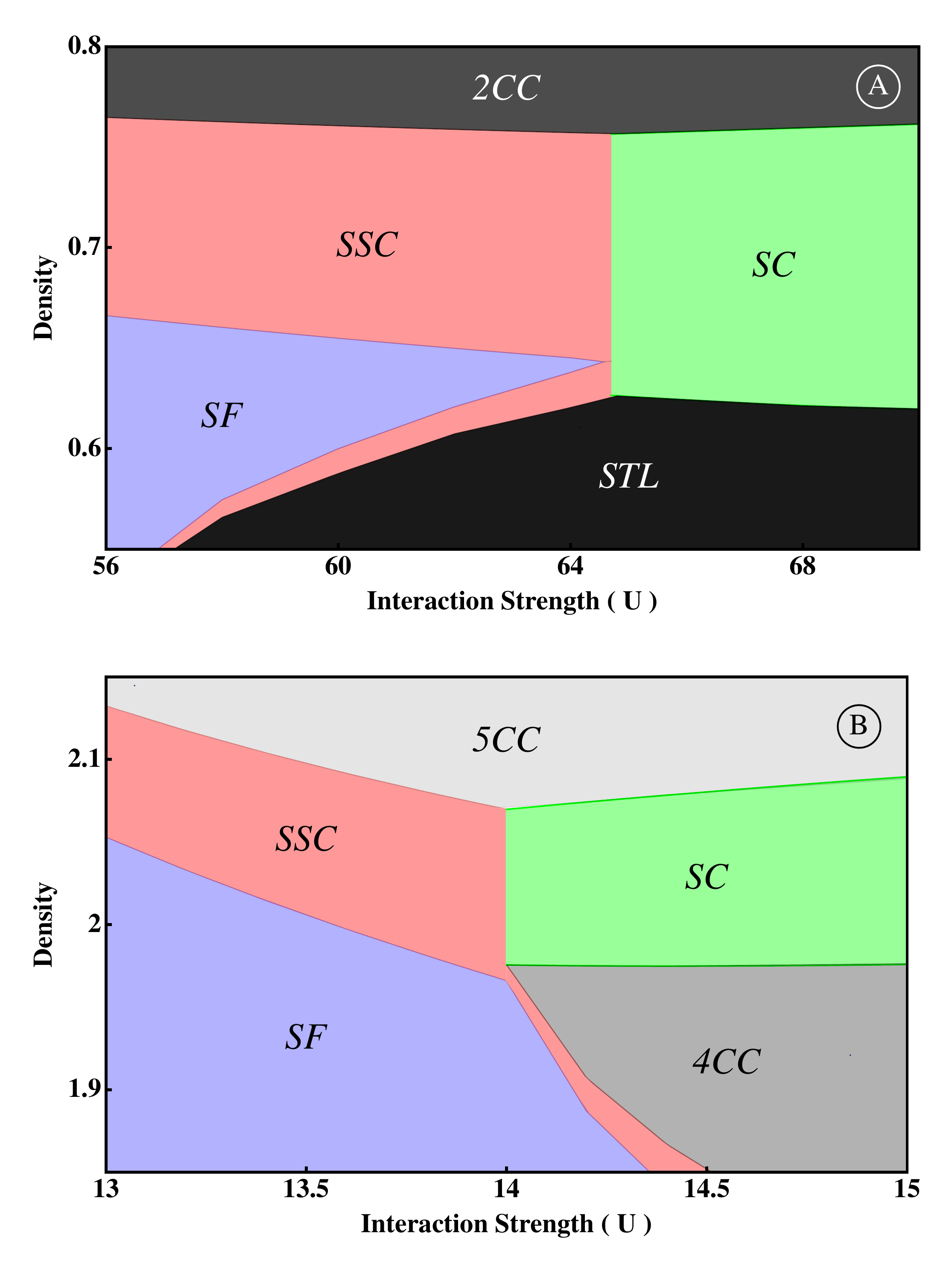}
	\caption{Zoom of the ground-state phase diagram of Fig.~\ref{fig2} remarking the coexistence regions of clusters with occupation $n$ and $n + 1$, and the superfluid.
	A) n = 1 and B) n = 7.
	Pure phases are represented in light blue for the superfluid region (SF), black for the single-particle triangular lattice (STL) and gray scale for the cluster crystal solids ($n$CC). 
	The coexistence regions are colored in green for the solid coexistence (SC) and orange for the solid-superfluid-coexistence (SSC).
	}
	\label{fig5}
\end{figure}

An interesting feature regarding the phase diagram is related to the boundary of the coexistence regions. 
From the thermodynamical point of view a system with only one component can not present an extended region in the phase diagram in which three phases coexist.
This constraint is a result of the Gibbs phases rule. 
At the same time, when the three coexistence regions solid($n$)-solid($n+1$), solid($n$)-superfluid and solid($n+1$)-superfluid meet there are a number of thermodynamical constraints that need to be fulfilled. 
This makes that the topology of the phase diagram, in the region where coexistence between different phases coincides, needs to reproduce the kind of behavior shown in Fig.~\ref{fig5}, independently of the level of approximation of the calculation.

To understand why this is a a general result we have to realize that, within each coexistence region, the chemical potential and the pressure have constant values along the density axis. 
At the same time, the pressure and the chemical potential need to be continuous functions of the density $\rho$ and of the strength of the interaction $U$. 
Consider now that at some value $U_0$ the coexistence region of the transition from a phase I to a phase II, increasing density, meets the coexistence region of the transition from the phase II to a phase III. 
This means that at $U_0$ the pressure and the chemical potentials of the pure I and III phases, at the boundary of the whole I-II-III coexistence region, are equals. 
And this is exactly the conditions for the establishment of the I-III coexistence region. 
The above arguments imply that at $U_0$ the energy curves resulting from the condition $\rho^2\frac{\partial \epsilon(\rho)}{\partial \rho}=\mathrm{constant}$, associated with the coexistence I-II, II-III and I-III, coincide. 
Consequently the overlap of the I-II and II-III coexistence regions with the I-III coexistence region have to be vertical, as shown in panels A and B of Fig.~\ref{fig5}.

It is worth noting, from Eq.~\ref{Urho} and Eq.~\ref{rhoco}, that for a given cluster state the extent in densities of the SSC region decreases as we approach to the bottom of the dome shape curve separating the solid from the superfluid phase (Fig.~\ref{fig4}). 
This mathematical result is a consequence of the fact that, at a given $U$, the angle between the curves $\epsilon_n(\rho)$ and $\epsilon_h(\rho)$ in its crossing point decreases to zero as we approach the limit of stability of the solid region. 
Lowering this angle implies that the curves are progressively more parallel and, consequently, the region in which we can further minimize the energy by creating a coexistence state, i.e. following an energy-density relation of the form  $\epsilon_c(\rho)=\epsilon(\rho_1)-P_c(\rho^{-1}-\rho_1^{-1})$,  is consistently diminished. 
The functionality $\epsilon_c(\rho)$ previously stated, is just a consequence of demanding that the pressure of the system within the coexistence region remains constant as the density is varied. 
This explanation clarifies why the extent of the SSC regions increases as we move away from the bottom of each solid dome.  

The above discussion clarifies the connection between the solid coexistence states (SC) with the SSC. 
In short, the coexistence regions between different pure cluster phases always evolve into a SSC phase as the interaction strength is decreased. 
It can be ensured that this transition occurs at a fixed value of $U$ in the whole density range of coexistence between different cluster phases and, at the moment of such transition, the extent of the SSC is the biggest possible. 
In this sense it can be understood that clusters coexistence favors the development of the SSC, as previous works have pointed out~\cite{cinti14}.

\section{Concluding remarks}

We have shown that the analytical understanding of the ground-state phase diagrams of ultrasoft bosons can be achieved by a variational approach when considering an integer number of particles per cluster.
With this assumption, the detailed shape of the phase diagrams can be improved with respect to mean-field calculations where a real number is considered for the cluster occupation.
The analytical approach introduced here is generally applicable to any ultrasoft potential. 

Among the determination of the regions of coexistence of solid phases, the solid-superfluid coexistence region (SSC) is particularly relevant. 
The comparison with previous Monte Carlo outcomes of bosonic systems revealed that the SSC region is visually coincident with the supersolid phase.   
The method can be used to find nontrivial information like detailed phase diagrams and possible markers for the emergence of supersolid regimes.

\section{Acknowledgements}
The work was supported by the Swedish Research Council Grants No. 642-2013-7837, 2016-06122, 2018-03659 and G\"{o}ran Gustafsson  Foundation  for  Research  in  Natural  Sciences  and  Medicine
and Olle Engkvists Stiftelse.
A.M.C. acknowledges financial support from Funda\c{c}\~ao de Amparo \`a Pesquisa de Santa Catarina, Brazil (Fapesc).


\bibliographystyle{apsrev4-1}
%

\end{document}